\documentclass[%
 reprint,
 amsmath,amssymb,
 aps,
prl,
]{revtex4-2}

\usepackage{graphicx}
\usepackage{dcolumn}
\usepackage{bm}

\usepackage{xcolor} 
\usepackage[normalem]{ulem} 

\usepackage{physics}

\begin{document}


%
\title{Interference-Based 3D Optical Cold Damping of a Levitated Nanoparticle}

\author{Youssef Ezzo}
\thanks{These authors contributed equally to this work.}
  
\author{Seyed Khalil Alavi\normalfont\textsuperscript{*}}%

\email{skalavi@fmq.uni-stuttgart.de}


\author{Sungkun Hong}
\email{sungkun.hong@fmq.uni-stuttgart.de}
\affiliation{
 Institute for Functional Matter and Quantum Technologies and\\ 
 Center for Integrated Quantum Science and Technology (IQST), University of Stuttgart,70569 Stuttgart, Germany \\
}%



\begin{abstract}
Achieving efficient three-dimensional feedback cooling of levitated nanoparticles is a key requirement for precision sensing and quantum control in levitated optomechanics. Here we demonstrate three-dimensional optical feedback cooling of a levitated nanoparticle using an interference-enhanced optical force generated within a single beam path. In this scheme, a weak auxiliary field co-propagates with the trapping tweezer and interferes with it to produce a tunable optical force that enables cold damping along all three center-of-mass motional axes without additional beam paths or trap reconfiguration. Using this approach, we cool a 142-nm-diameter silica nanoparticle in high vacuum to effective temperatures of 625.8, 711.6, and 19.9 mK along the $x$, $y$, and $z$ directions, respectively, at a pressure of $8.5\times10^{-6}$ mbar. The cooling dynamics and their dependence on feedback gain and pressure are well described by a cold-damping model. Because the feedback force is generated optically, the scheme does not rely on electrical actuation and is directly compatible with neutral particles. These results establish interference-based optical forces as a simple and broadly applicable mechanism for three-dimensional feedback control in levitated optomechanics, with a clear pathway toward the quantum regime under improved vacuum and detection conditions.
\end{abstract}

\maketitle



\section{Introduction}

Levitated optomechanics has emerged as a powerful platform to explore quantum behavior on the mesoscopic mass scale~\cite{gonzalez-ballestero_levitodynamics_2021} and to develop ultra-sensitive force and acceleration sensors~\cite{ranjit_zeptonewton_2016, monteiro_force_2020, moore_searching_2021}. Nanoparticles trapped by optical tweezers in high vacuum benefit from exceptional environmental isolation, allowing observation of quantum coherent motions even at room temperature ~\cite{chang_cavity_2010, romero-isart_toward_2010}. A central requirement for these applications is efficient cooling of the particle's center-of-mass (CoM) motion, to stabilize the particle in high vacuum and to enable cooling toward the motional ground state.

Using cavity-assisted schemes, researchers have cooled motions of levitated particles to the quantum ground state, including both translational~\cite{delic_cooling_2020, ranfagni_two-dimensional_2022, piotrowski_simultaneous_2023} and librational modes~\cite{dania_high-purity_2025, troyer_quantum_2025_2}. 
As a versatile cavity-free alternative, measurement-based feedback control has also been widely implemented, notably through electrically generated cold  damping~\cite{tebbenjohanns_cold_2019, conangla_optimal_2019, kremer_all-electrical_2024}. This approach has also enabled ground-state cooling of levitated nanoparticles~\cite{magrini_real-time_2021, tebbenjohanns_quantum_2021}.
However, the net charge required in such schemes can introduce excess decoherence and technical noise~\cite{hebestreit_sensing_2018, kamba_quantum_2025}, motivating the development of purely optical feedback methods compatible with neutral particles.

Several optical cold-damping techniques have been demonstrated, including schemes based on scattering forces from single \cite{dania_optical_2021} or counter-propagating beams~\cite{li_millikelvin_2011}, fast steering of the trapping beam~\cite{vijayan_scalable_2023}, and frequency modulation of standing-wave traps~\cite{kamba_optical_2022}, the latter recently achieving ground-state cooling. While these approaches have achieved remarkable performance, they typically rely on multiple optical paths or specific beam geometries and are often restricted to cooling only selected motional directions. Such constraints can limit their flexibility, scalability, or simultaneous cooling of all motions.

Here, we present an interference-based optical cold damping technique that enables simultaneous cooling of all three CoM motional degrees of freedom using a single co-propagating auxiliary beam. The method relies on controlled interference between a primary optical tweezer and a weak auxiliary beam, generating a fully tunable optical force acting on the nanoparticle~\cite{alavi_interference-enhanced_2025}. Using this method, we demonstrate three-dimensional cooling of a levitated nanoparticle to temperatures of 625.8, 711.6, and 19.9~mK along the $x$, $y$, and $z$ axes, respectively, at a pressure of $8.5\times10^{-6}$~mbar. The scheme is simple to implement and does not require multiple beam paths, providing a scalable route toward fully optical feedback control of neutral nanoparticles.

\section{Experimental setup}

\begin{figure*}
    \centering
    \includegraphics{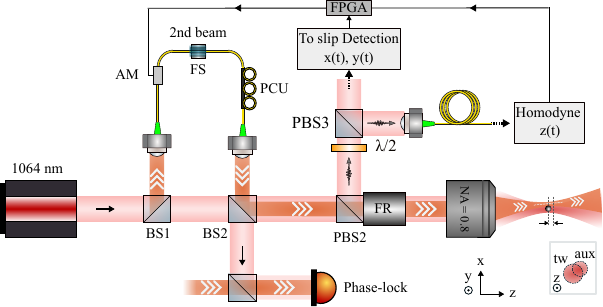}
    \caption{
    Experimental setup. A silica nanoparticle is trapped in high vacuum by a 1064~nm optical tweezer formed by NA$=0.8$ objective. Backscattered light is collected by the same objective and routed via a Faraday rotator (FR) and a polarizing beam splitter (PBS) to the detection path, where balanced homodyne detection measures the axial ($z$) motion and split photodetectors measure the transverse ($x,y$) motions. An auxiliary beam derived at BS1 passes through a fiber-coupled amplitude modulator (AM) and a fiber stretcher (FS), and is then recombined with the trapping beam at BS2 to co-propagate to the trap. A slight displacement between the two beams gives rise to an interference-induced gradient force with nonzero components along all three spatial directions (inset). Feedback is implemented by electronically processing the measured motional signals with appropriate gains and delays and applying the resulting drive to the AM.
}
    \label{fig: setup}
\end{figure*}

As illustrated in Fig.~\ref{fig: setup}, we trap a silica nanosphere (nominal diameter 142~nm) in high vacuum using an optical tweezer formed by tightly focusing a 1064~nm laser with an NA~$=0.8$ microscope objective. With 410~mW of optical power at the focus, the particle’s CoM oscillation frequencies are $(\Omega_x,\Omega_y,\Omega_z)/2\pi \approx (315,\, 284,\, 83)$~kHz.

The CoM motion is read out from the light backscattered by the particle. The backscattered field is collected by the same objective and routed to the detection optics via a Faraday rotator (FR) and a polarizing beam splitter (PBS). The detection path is split into two channels: a transverse readout that yields $x(t)$ and $y(t)$ via the split-detection scheme, and an axial readout in which the $z(t)$ is measured by balanced homodyne detection. These signals are processed electronically and used to generate optical feedback forces.

In our implementation, the feedback force $F_{\mathrm{fb}}$ is realized via coherent interference between the primary tweezer field and a co-propagating weak auxiliary field~\cite{alavi_interference-enhanced_2025}. Their interference imposes an additional phase-dependent term on the optical potential, giving rise to a tunable force on top of the nominal tweezer forces. By introducing a small relative offset between the focal positions of the two beams, the force at the particle location can be engineered to have nonzero projections along $x$, $y$, and $z$ (see Supplemental Material~\cite{SM}). Its amplitude scales as $F_{\mathrm{fb}} \propto \sqrt{P_{\mathrm{tw}}\cdot P_{\mathrm{aux}}}$, where $P_{\mathrm{tw}}$ and $P_{\mathrm{aux}}$ denote the optical powers of the tweezer and auxiliary beams, respectively~\cite{alavi_interference-enhanced_2025}. Feedback cooling is implemented by modulating $P_{\mathrm{aux}}$ in response to the measured motion such that the resulting force is approximately proportional to the particle velocity (see Ref.~\cite{SM}), thereby providing damping of all CoM modes.

Experimentally, the auxiliary beam is derived from the main 1064~nm laser at BS1, which serves as an effective pick-off stage, and is sent through a fiber-coupled amplitude modulator (AM), which provides fast control of $P_{\mathrm{aux}}$. The measured signals $x(t)$, $y(t)$, and $z(t)$ are electronically filtered and phase-shifted by the desired delays, and the resulting drive signals are applied to the AM to realize the feedback force $F_{\mathrm{fb}}(t)$.

A fiber stretcher (FS) in the auxiliary arm adjusts the optical path length and hence the phase of the auxiliary beam with respect to the primary tweezer field. We actively lock this phase by interfering a small fraction of the main field from the unused output port of BS2 with a power-matched portion of the auxiliary beam; the resulting error signal is fed back to the FS, achieving a fixed interference condition throughout the experiment.

\section{Experimental results}
\begin{figure}[h]
    \centering
    \includegraphics{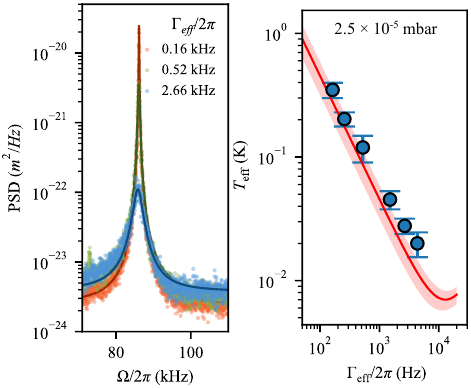}
    \caption{(a) Power spectral densities (PSDs) of the particle motion along the $z$ axis, measured at a pressure of $2.3\times10^{-5}$~mbar for different feedback gains, corresponding to effective damping rates $\Gamma_{\mathrm{eff}} \approx 0.16$~kHz (orange), $0.52$~kHz (green), and $2.66$~kHz (blue). Solid lines are fits to Eq.~\ref{eq:motion_fft_IL_PSD}. (b) Extracted effective temperature $T_{\mathrm{eff}}$ of the $z$-axis motion as a function of $\Gamma_{\mathrm{eff}}$. 
    Each data point represents the mean of ten independent measurements (1~s acquisition time each), with error bars indicating the standard deviation. 
    The solid line shows the theoretical prediction based on Eq.~(\ref{eq:Teff}), while the shaded region indicates the uncertainty due to the imprecision of the pressure gauge ($\approx 30\%$).}    
    
    \label{fig: gainChange}
\end{figure}

The center-of-mass (CoM) motion of the trapped nanoparticle along each spatial direction $q \in \{x,y,z\}$ can be modeled as that of a damped harmonic oscillator subject to thermal noise and an externally applied feedback force. Its dynamics are described by the Langevin equation
\begin{equation}
    m \ddot{q}(t) + m \Gamma_g \dot{q}(t) + m \Omega_0^2 q(t)= F_{\mathrm{th}}(t) + F_{\mathrm{fb}}(t),
    \label{eq:langevin}
\end{equation}
where $m$ is the particle mass, $\Omega_0$ is the mechanical resonance frequency, and $\Gamma_g$ is the damping rate due to residual gas collisions. The thermal force $F_{\mathrm{th}}(t)$ has zero mean and satisfies $\langle F_{\mathrm{th}}(t)F_{\mathrm{th}}(t') \rangle = 2 m \Gamma_g k_B T\,\delta(t-t')$, where $T$ is the bath temperature. 
The cold damping scheme is realized by applying a feedback force proportional to the measured particle displacement and delayed appropriately, such that $F_{\mathrm{fb}}(t)\approx -m\Gamma_{\mathrm{fb}}\dot{q}(t)$ \cite{tebbenjohanns_cold_2019, conangla_optimal_2019, SM}. Here, $\Gamma_{\mathrm{fb}}$ denotes the feedback-induced damping rate.
This force increases the total effective damping to $\Gamma_{\mathrm{eff}} = \Gamma_g + \Gamma_{\mathrm{fb}}$, resulting in a reduction of the effective temperature of the CoM motion.

Following the cold damping scheme with our interference-based optical force, we demonstrate cooling of the particle's motion. Figure~\ref{fig: gainChange}(a) shows the measured power spectral densities (PSD) of the $z$-axis motion for different feedback gains at a pressure of $2.3\times10^{-5}$~mbar. We note that, during the experiment, we also apply cooling of the transverse ($x,y$) motions with fixed gains to suppress any nonlinear inter-modal couplings \cite{magrini_real-time_2021, tebbenjohanns_quantum_2021}. For each gain setting, the PSD is obtained by averaging ten consecutive time traces of 1~s duration. The effective temperature is extracted by fitting the measured PSDs to the standard in-loop detection model~\cite{conangla_optimal_2019, melo_vacuum_2024, vijayan_scalable_2023}, which accounts for feedback-induced correlation between detector noise and the particle (see Ref.~\cite{SM}):
\begin{equation}
S^{\mathrm{IL}}x(\Omega) =\frac{(2k_B T \Gamma_g)/m^2}{(\Omega_0^2 -\Omega^2 )^2 + \Omega^2 \Gamma_{\text{eff}} ^2}+\frac{(\Omega_0^2 -\Omega^2 )^2 + \Omega^2 \Gamma_g^2}{(\Omega_0^2 -\Omega^2 )^2 + \Omega^2 \Gamma_{\text{eff}}^2}\sigma_N^2 ,
\label{eq:motion_fft_IL_PSD}
\end{equation}
where $\sigma_N^2$ denotes the displacement measurement noise floor. From the fitting results, the effective temperature $T_{\mathrm{eff}}$ is deduced (see Ref.~\cite{SM}):
\begin{equation}
T_{\mathrm{eff}} =
T\frac{\Gamma_g}{\Gamma_{\mathrm{eff}}}
+ \frac{m\Omega_0^2 \Gamma_{\mathrm{fb}}^2}{k_B \Gamma_{\mathrm{eff}}}\sigma_N^2 .
\label{eq:Teff}
\end{equation}
The first term in Eq.~(\ref{eq:Teff}) corresponds to the familiar cold-damping result \cite{cohadon_cooling_1999, poggio_feedback_2007}, where the effective temperature is reduced proportionally to the ratio $\Gamma_g/\Gamma_{\mathrm{eff}}$. The second term arises from measurement noise fed back into the system and sets the ultimate limit to feedback cooling.

The extracted effective temperatures are shown in Fig.~\ref{fig: gainChange}(b). Each data point denotes the mean $T_{\mathrm{eff}}$, with error bars indicating the standard deviation. As expected, $T_{\mathrm{eff}}$ decreases with increasing feedback-induced damping rate $\Gamma_{\mathrm{fb}}$. The solid curve in Fig.~\ref{fig: gainChange}(b) is the the prediction of Eq.~\ref{eq:Teff}, evaluated for $T=300$~K with $\Gamma_g$ estimated from kinetic theory in the rarefied-gas regime~\cite{gas_damping_theory_1990}. The excellent agreement between experiment and theory confirms that our cooling performance is well captured by the cold-damping model.

\begin{figure}
    \centering
    \includegraphics{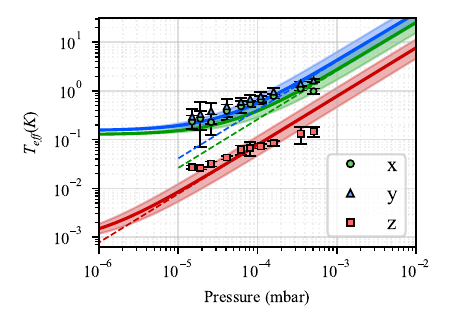}
    \caption{Extracted effective temperatures $T_{\mathrm{eff}}$ of the particle motion along the $x$ (red), $y$ (green), and $z$ (blue) axes as a function of background pressure. Each data point represents the mean temperature extracted from 12 consecutive PSDs, each calculated from a 50 ms time trace, with error bars indicating the standard deviation. As the pressure is reduced, $T_{\mathrm{eff}}$ decreases linearly (dashed lines), consistent with the expected pressure scaling of the gas damping rate. At lower pressures, deviations from linearity appear and $T_{\mathrm{eff}}$ saturates as the contribution of measurement noise (second term in Eq.~(\ref{eq:Teff})) becomes significant. Solid lines show the theoretical prediction based on Eq.~(\ref{eq:Teff}), and the shaded regions indicate the uncertainty due to the pressure gauge imprecision.}
    
    \label{fig: TeffPressure}
\end{figure}

The effective temperature also depends on the background gas pressure, as the gas damping rate scales linearly with pressure, $\Gamma_g \propto p$. This behavior is summarized in Fig.~\ref{fig: TeffPressure}, which shows the extracted effective temperatures as a function of pressure for the $x$, $y$, and $z$ motions. For these measurements, the feedback gain is kept fixed as the pressure is varied. At higher effective temperatures, $T_{\mathrm{eff}}$ scales linearly with pressure, consistent with the expected pressure dependence of $\Gamma_g$. At lower effective temperatures, deviations from linearity, most notably for the transverse modes, appear as measurement noise becomes relevant and ultimately limits further cooling.

\begin{figure}[h]
    \centering
    \includegraphics{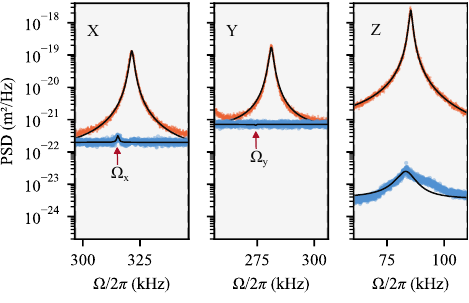}
    \caption{PSDs of the particle motion along the $x$, $y$, and $z$ axes, measured using split detections for the transverse directions ($x$, $y$) and homodyne detection for the axial direction ($z$). Each panel compares the PSDs measured without feedback cooling at a pressure of 2~mbar (red) and with simultaneous cooling of all three motions activated at a pressure of $8.6\times10^{-6}$~mbar (blue). Feedback cooling strongly suppresses the mechanical resonances, leading to effective temperatures of $(625.8,\, 711.6,\, 19.9)$~mK for the $x$, $y$, and $z$ motions, respectively. For the transverse directions, the cooled resonances approach the detection noise floor. The resonance frequencies are indicated by arrows.}
    \label{fig: psd_cooled}
\end{figure}

At a pressure of $8.5\times10^{-6}$~mbar, we reach minimum effective temperatures of (625.8,\, 711.6,\, 19.9)~mK for the motions along $x$, $y$, and $z$, respectively. Figure~\ref{fig: psd_cooled} shows the corresponding PSDs of the cooled particle motions along all three axes. PSDs recorded without feedback at a pressure of 2~mbar are also shown for comparison. Along the axial direction, the cooled signal remains well above the detection noise floor ($\sim 1.9~\mathrm{pm}/\sqrt{\mathrm{Hz}}$), indicating that the cooling is predominantly limited by residual gas collisions rather than measurement noise. In contrast, the cooled transverse signals approach the detection limits of $\sim 14.3$ and $\sim 26.5~\mathrm{pm}/\sqrt{\mathrm{Hz}}$ for the $x$ and $y$ axes, respectively, leaving only barely discernible resonance peaks.

\section{Discussion and conclusion}

In conclusion, we have demonstrated an interference-based optical cold-damping scheme using a co-propagating auxiliary beam, which enables full three-dimensional cooling of a levitated nanoparticle. The method is simple to implement, does not rely on multiple beam paths or complex trap geometry. It is therefore readily compatible with a broad range of levitated optomechanical platforms, including microcavities~\cite{magrini_near-field_2018, alavi_enhanced_2025}, near-surface trapping architectures~\cite{diehl_optical_2018, ju_near-field_2023}, and scalable multiparticle configurations~\cite{vijayan_scalable_2023}. These results establish interference-induced optical forces as a versatile tool for cooling and control in levitated optomechanics.

The prospect of ground-state cooling in the present scheme is currently limited by different mechanisms along different motional axes. Along the axial ($z$) direction, the minimum temperature is set by residual gas collisions, as evidenced by the observed linear pressure scaling of $T_{\mathrm{eff}}$. Extrapolating this behavior indicates that the ground-state level, with mean occupation $\bar{n} = k_B T_{\mathrm{eff}}/\hbar\Omega_0\lesssim 1$, would become accessible at pressures on the order of $1 \times 10^{-8}$~mbar~\cite{magrini_real-time_2021,kamba_optical_2022}, where the recoil heating dominates the decoherence process.

In contrast, the transverse ($x,y$) modes are already limited by detection noise at the lowest achieved pressures. In the present detection geometry, the displacement information for transverse motion is intrinsically limited, since the scattered field carries maximal position information in directions perpendicular to the optical axis~\cite{tebbenjohanns_optimal_2019}. Collecting scattered light in a transverse direction, for example, using high-NA fiber-coupled microlenses~\cite{alavi_compact_2026}, would substantially enhance the transverse displacement sensitivity and enable further cooling of the transverse modes.

\begin{acknowledgments}
\subsection{Acknowledgments}
This research was funded by Deutsche Forschungsgemeinschaft (Projektnummers: 523178467) and Carl-Zeiss-Stiftung (Johannes-Kepler Grant through IQST).

\subsection{Disclosures}
The authors declare no conflicts of interest.

\subsection{Data availability}

The data that support the findings of this article are not publicly available. The data are available from the author upon reasonable request.

\end{acknowledgments}

\bibliography{refs}

\end{document}